\documentclass[runningheads]{llncs}

\let\llncssubparagraph\subparagraph
\let\subparagraph\paragraph
\usepackage[compact]{titlesec}
\let\subparagraph\llncssubparagraph

\usepackage{graphicx}
\usepackage{comment}
\usepackage{times} 
\usepackage{url} 
\usepackage{graphicx} 
\usepackage{authblk} 
\usepackage{amsmath} 
\usepackage{amsfonts}
\usepackage{stmaryrd}
\usepackage{tabularx,ragged2e,booktabs} 
\usepackage{titlesec}
\usepackage{amssymb}
\usepackage{comment}
\usepackage{subcaption}
\usepackage{authblk}
\usepackage{xcolor} 
\usepackage{hyperref}
\usepackage{amssymb}
\usepackage{dsfont}
\usepackage{float}
\usepackage[ruled,vlined]{algorithm2e}
\begin{document}
\title{A Joint Graph and Image Convolution Network for Automatic Brain Tumor Segmentation.}
\titlerunning{Joint GNN-CNN for brain tumor segmentation}
%
\author{Camillo Saueressig \inst{1,2}\orcidID{0000-0001-6372-0695} \and
Adam Berkley \inst{1} \and
Reshma Munbodh \inst{3,*}\orcidID{0000-0002-7982-7814} \and
Ritambhara Singh \inst{1,2,*}\orcidID{0000-0002-7523-160X}}
\authorrunning{C. Saueressig \textit{et al.}}
%
\institute{Department of Computer Science, Brown University \and
Center for Computational Molecular Biology, Brown University \and
Department of Radiation Oncology, Brown Alpert Medical School 
 *\email{\{reshma\_munbodh,ritambhara\}@brown.edu}}
\maketitle              
\begin{abstract}


We present a joint graph convolution – image convolution neural network as our submission to the Brain Tumor Segmentation (BraTS) 2021 challenge. We model each brain as a graph composed of distinct image regions, which is initially segmented by a graph neural network (GNN). Subsequently, the tumorous volume identified by the GNN is further refined by a simple (voxel) convolutional neural network (CNN), which produces the final segmentation. This approach captures both global brain feature interactions via the graphical representation \textit{and} local image details through the use of convolutional filters. We find that the GNN component by itself can effectively identify and segment the brain tumors. The addition of the CNN further improves the median performance of the model on the validation set by 2 percent across all metrics evaluated.

\end{abstract}

\section{Introduction}

\quad Tumor segmentation is a cornerstone of nearly all standard tumor treatments. It is integral for surgical and radiation planning, treatment response analysis, and longitudinal tumor monitoring, among other standard practices. However, manual tumor segmentation is notoriously time-consuming and subjective, even for highly trained radiologists. Automatic tumor segmentation can produce such segmentations in a fraction of the time in a standardized, reproducible fashion.
Over the past decade, the performance of automated biomedical segmentation methods has significantly improved across multiple tumor types, and brain tumors are no exception~\cite{haque2020deep,bi2019artificial}. The Brain Tumor Segmentation dataset (BraTS) is the largest publicly available dataset of glioma MRIs and corresponding expert segmentations and has played a pivotal role in developing and evaluating these methods~\cite{menze2014multimodal,bakas2017advancing,bakas2018identifying,bakas2017segmentation1,bakas2017segmentation2}.

The 2021 BraTS tumor segmentation challenge consists of over 2000 multi-para- metric magnetic resonance images (MRIs) of tumorous brain volumes (specifically, gliomas) imaged across a wide array of institutions. While the images are compiled from a number of different institutions, they are all processed using a standard pipeline, and the same four modalities are available for every volume.
These are T1-weighted, T1-weighted contrast-enhanced, T2-weighted, and Fluid Attenuated Inversion Recovery (FLAIR) modalities, all of which provide complementary information on the location and shape of the tumor and its compartments. The ground truth labels are generated using an ensemble of top-performing models from previous years and are manually revised by an expert neuroradiologist for all images. The challenge aims to correctly classify each voxel of a given brain volume as either healthy tissue, edema, enhancing tumor (ET), or necrotic tumor core. These tumor sub-regions can be combined into the whole tumor (WT) and core tumor (necrotic core + enhancing tumor, CT) to further evaluate model performance on gross tumor segmentation~\cite{baid2021rsna}.

\begin{figure}[ht]
\includegraphics[width=\textwidth]{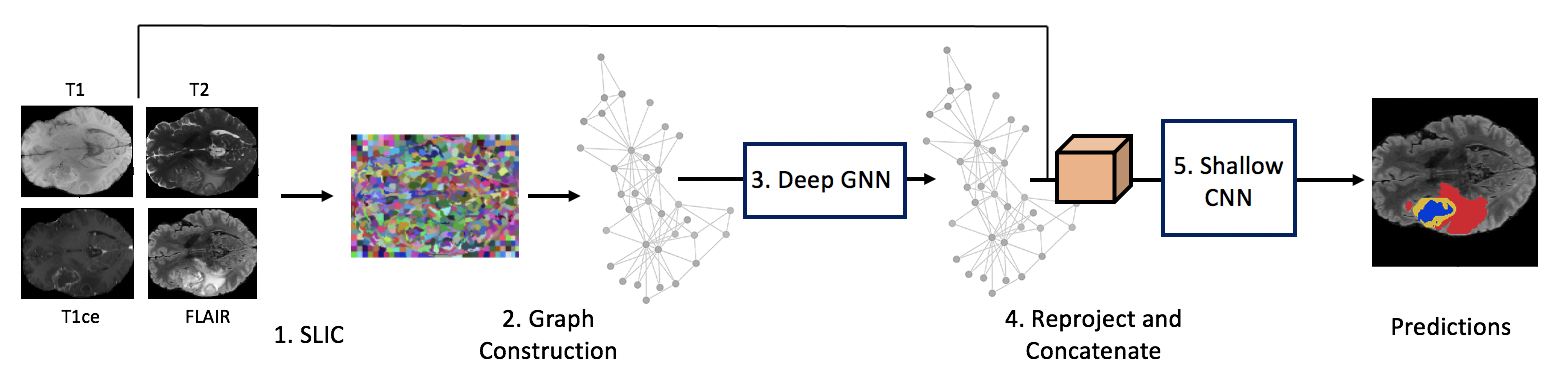}
\caption{\textbf{GNN-CNN Model Overview.} MRI Modalities are first stacked to create one 3D Image with 4 channels. 1) Combined modalities are clustered into supervoxels using SLIC. 2) Supervoxels are converted to a graph structure such that each supervoxel becomes one graph node (depicted graph is greatly simplified). 3) Graph is fed through a Graph Neural Network 4) Node prediction outputs (more specifically, logits) are overlaid back onto the supervoxels. The original input image features are concatenated with re-projected node logits. 5) The result is fed through a 2-layer CNN which produces final predictions.} \label{model_overview}
\end{figure}

Our submission to the BraTS 2021 challenge is a joint graph neural network (GNN) - convolutional neural network (CNN) model (summarized in Figure \ref{model_overview}). The GNN module aims to partition the brain into distinct regions and predict the label of each region, and the CNN component refines the predictions made by the GNN. Unlike the vast majority of BraTS competitors in recent years~\cite{bakas2018identifying}, which exclusively perform inference directly on voxel data, our model instead learns and predicts primarily on a graphical representation of the brain. We model each brain volume as composed of small, contiguous regions and connect nearby regions using edges, forming a graph. Each graph node contains information summarizing the intensity information of the brain in that region across all four modalities, and the edges allow neighboring regions to share their information with each other. This formulation greatly simplifies the representation of a brain from millions of voxels down to only thousands of nodes, while preserving nearly all the information. It also enables the modeling of explicit connectivity between different regions of the brain and potential long-range interactions between distant regions, which are difficult to capture using only CNNs. We have previously developed a similar model composed only of a graph neural network on the 2019 BraTS dataset~\cite{ourpaper}. Here, we improve on our previous work by adding a shallow CNN to the end of the model, which smooths out the model predictions at region boundaries and provides a substantial ($\geq 2\%$) improvement in both median Dice score and median Hausdorff distance on the validation set.

\section{Methods}

Our GNN-CNN model is composed of two components. The core component is a graph neural network (GNN)~\cite{kipf2016semi,zhou2018graph}. For a given input graph representing one patient sample, where each node corresponds to a collection of adjacent voxels in the original MRI image, the GNN predicts each node's label. Since the GNN can only predict the label of nodes (i.e. brain regions) atomically, its predictions are necessarily coarser than voxel-based predictions. This property can lead to incorrect predictions at the edges of tumor compartments, where created regions can contain voxels of multiple labels~\cite{ourpaper}. This shortcoming is especially pronounced in small tumors. Accordingly, we have added a second component to our model: a shallow CNN~\cite{lecun2015deep}. The convolutional layers receive both the GNN prediction logits (projected back into an image) and the original voxel image data. They are thus able to make fine-grained adjustments to the coarse predictions based on local voxel information. The details of the model are presented in Figure~\ref{model_detailed}.

\subsection{Graph Construction from MRI Modalities}
Both the input and the output of the GNN are required to be graph-structured data. Therefore, before feeding the MRI scans into our network, we transform them into graphs. Graphs are composed of nodes and edges, where both the nodes and the edges can have features associated with them. In this work, each node corresponds to one image region, and an edge between two nodes corresponds to spatial proximity of the corresponding regions. We partition the brain into regions using supervoxels. Supervoxels are the 3D analog to superpixels, i.e., collections of nearby pixels that share similar intensities. 

We construct the supervoxels using the Simple Linear Iterative Clustering (SLIC) algorithm~\cite{achanta2012slic}. SLIC uses a combination of spatial and intensity information to partition an image into approximately a desired number of supervoxels using K-means clustering. While the input to SLIC is traditionally in either RGB or Lab color space, we find that running SLIC directly on the stacked MRI modalities still produces meaningful supervoxels. To determine the optimal hyperparameters for the SLIC algorithm, we perform a grid search across \textit{k}, the number of supervoxels and \textit{m}, the compactness coefficient (the weighting between spatial and intensity information), and compute the achievable segmentation accuracy (ASA). ASA measures how well the GNN would perform on a given supervoxel partitioning, given that it classifies every supervoxel according to the most common label of the constituent voxels. The ASA is high if there is a strong correspondence between supervoxel shape and tumor boundaries, resulting in supervoxels composed of voxels with the same label. It is low if supervoxels are composed of voxels with mixed labels.

After the supervoxels are generated via SLIC, we discard those supervoxels that lie outside the brain volume. Of the remaining supervoxels, each is assigned a feature vector, a label, and a set of neighbors. The feature vector summarizes the intensities of the input MRIs for its comprising voxels. We empirically found that intensity quintiles for each modality yielded the best results. The label is the majority label (mode) of its constituent voxels. The neighbors of a supervoxel are all other supervoxels which are directly adjacent to it. A graph is then constructed where each supervoxel forms one node with its associated features and label, and each supervoxel shares an unweighted and undirected edge with its neighbors.

\subsection{GNN Architecture}
\begin{figure}[ht]
\includegraphics[width=\textwidth]{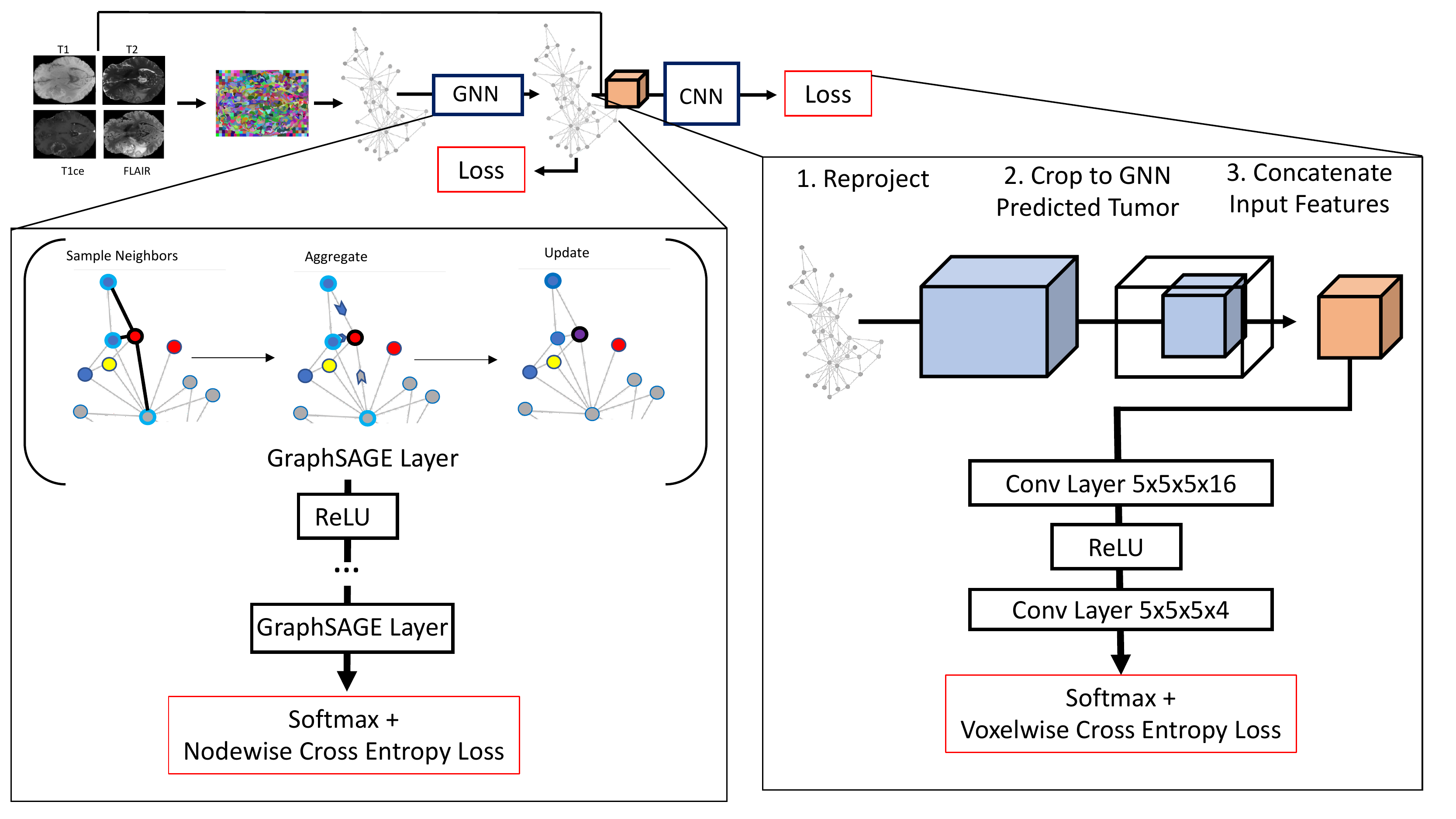}
\caption{\textbf{Detailed view of GNN and CNN}. \textbf{Left}: The GNN is composed of GraphSAGE layers alternated with a nonlinearity. Each GraphSAGE layer updates each node's features by sampling neighboring nodes and aggregating the features (Eq.\ref{eq:gs-pool}). \textbf{Right}: 1) The output of the GNN is reprojected into a 3D image by assigning each voxel the output logits of its corresponding node. 2) Based on this reprojection, the approximate location of the tumor predicted by the GNN is located and cropped out. 3) The projected and cropped logits are concatenated with the image features for that same location. This volume is then fed through a two-layer CNN. Note that the output of both the GNN and CNN components have an associated loss function. } \label{model_detailed}
\end{figure}

Our graph neural network is composed of several sequential GraphSAGE-pool layers~\cite{hamilton2017inductive} alternated with the ReLU non-linearity (Figure \ref{model_detailed}). Each layer transforms the features of each node by aggregating information from that node's neighbors, according to Eq.~\ref{eq:gs-pool}
\begin{equation}
h^{(l+1)}_u = \sigma(W^{(l)} \cdot (h^{(l)}_u \ || \  max(\sigma(W_{pool} \cdot h^{(l)}_v) \ \forall \ v \in V(u)))
\label{eq:gs-pool}
\end{equation}
where $h_u^{(l)}$ is the features of node $u$ at layer $l$, $\sigma$ is a differentiable, non-linear activation function, $W^{(l)}$ is a layer specific trainable weight matrix, $W_{pool}$ is a global trainable weight matrix, $||$ is the concatenation operator, and $V(u)$ is the subset of nodes which are are directly connected to $u$ via edges, also known as the neighborhood of $u$.

The input layer expects 20 features (5 quintiles for each of four modalities) and the output layer outputs 4 logits (one for each label). The output logits are duplicated, where one copy is passed directly through a loss function which backpropagates only through the GNN, and the other is passed through to the CNN (Fig. ~\ref{model_detailed})

\subsection{CNN Architecture}

The CNN consists of two convolutional layers with a $5\times5\times5$ kernel size and a stride of 1 (Figure ~\ref{model_detailed}). The first layer has $16$ filters and the second $4$ (one for each label) with ReLU nonlinearity between the two layers. The architecture is purposefully kept simple since it only serves to refine the predictions made by the GNN.

The input to the CNN is the concatenation of the GNN output logits ($f=4$) and the input MRI modalities ($f=4$) for each voxel. Therefore, the CNN receives the predictions of the GNN in addition to the image features, which allows it to correct the predictions made by the GNN. This correction is especially relevant around the edges of the tumor and its compartments, where the coarse predictions from the GNN can often result in misclassifications of strips of voxels. We feed only the tumorous tissue through the CNN to reduce the memory requirement and computation time. Specifically, we crop out a patch of the volume containing the tumor, as predicted by the GNN, and the CNN further refines only that patch.

\subsection{Loss Functions}
We calculate and backpropagate loss through our model at two locations. A voxel-wise cross-entropy loss is calculated from the output of the CNN and backpropagated \textit{only} through the convolutional layers. This loss is unweighted as the input to the CNN has been cropped to the tumor-containing volume. 

A node-wise weighted cross-entropy loss is calculated from the GNN logits and backpropagated through the GNN. The ground truth label for each node is generated by finding the mode of the labels in the corresponding supervoxel. This loss is weighted approximately inversely to the prevalence of each label to address the class imbalance.

We include this GNN loss function to obtain prediction logits of the nodes that can then be easily projected in the image space. It is crucial for the model's performance that the GNN output be interpretable as predictions, so that the predicted tumorous volume can be located and cropped out. Furthermore, this formulation allows us to visualize the finer corrections that the CNN layer performs over the coarse GNN predictions (see Fig.~\ref{ex_results_val} for example).

\subsection{Model Training}

In practice, we train the GNN and CNN sequentially rather than simultaneously to decrease training time. The GNN is trained for 300 epochs on mini-batches of 6 graphs, whereas the CNN is trained for 100 epochs using only one sample at a time. The training of a full model takes approximately 2 days on an 8GB GPU.

We used the AdamW optimizer with weight decay of 0.0001 and exponentially decrease learning rate according to Eq.~\ref{eq:exp-decay}
\begin{equation}
lr_{e}=lr_{0}*\lambda^{e}
\label{eq:exp-decay}
\end{equation}
where $lr_{0}$ is the initial learning rate, $e$ is the current epoch and $\lambda=0.98$. We found that adding additional regularization, such as dropout or higher weight decay, did not improve performance.

The BraTS 2021 dataset is split into training(n=1251), validation(n=216), and test(n=570) partitions. The hyperparameters for only the GNN component, i.e., GNN layer sizes, GNN depth, learning rate, and class weighting, were tuned using random search and 5-fold cross-validation on the entire training set (n=1251). The GNN architecture with the best average performance across the 5 folds was then integrated into the full hybrid model. Three architectural replicates were trained on the entire dataset and evaluated on the validation set. The best performing replicate was then submitted for evaluation on the test dataset. We report the mean and median results of the best performing replicate on both the validation and test sets in Section ~\ref{sec:performance}

\subsection{Data Preprocessing}
The BraTS dataset MRIs are all padded to a standard shape to facilitate image-based processing. Since our approach is primarily graph-based and does not rely on uniform input sizes, we first crop each patient sample to the tightest possible bounding box around the brain to minimize the amount of background volume prior to supervoxel creation. Subsequently, we rescale each MRI to the approximate $[0,1]$ range by dividing by the 99.5 percentile of intensity values in that MRI. The raw MRI data is not collected in a bounded range and can vary by several orders of magnitude even between two images of the same modality. As such, this step normalizes the intensity values to be consistent across the dataset. Finally, we compute the mean and standard deviation for each modality across the entire training dataset (on non-zero voxels) and standardize each modality to have zero mean and unit variance.

\section{Results}

\subsection{Hyperparameters}

The SLIC parameters with the highest achievable segmentation accuracy (ASA) were $k=15000$ and $m=0.5$. The value for $m$ differs from that in our previous work~\cite{ourpaper} as our preprocessing steps have slightly changed.

The best performing GNN model from the cross-validation phase had 6 layers with 256 neurons each and a learning rate of 0.0005. The GNN is thus deeper and has many more learnable parameters than the CNN. This is a purposeful design choice to force the GNN to do the majority of the learning.

\subsection{Evaluation Metrics}

The performance of the models submitted to the BraTS challenge are evaluated using two metrics, Dice score and the $95^{th}$ percentile of the symmetric Hausdorff distance. Both metrics are evaluated over the whole tumor, core tumor, and active tumor subregions. Intuitively, the Dice score measures the overlap between the predictions and the ground truth while Hausdorff distance measures the most the predicted and ground truth segmentations diverge from each other.

\begin{equation}
 \text{Dice} = \frac{2TP}{2TP + FP + FN}.\label{eq:dice_eqn}
\end{equation}
 where $TP$, $FP$, and $FN$ are the number of true positives, false positives, and false negatives, respectively. True positive voxels are defined as those correctly assigned as belonging to a specific tumor compartment.

\begin{equation}
 \text{HD95} = 95\% \ (d(\hat{Y},Y) || d(Y,\hat{Y})) \label{eq:hd95}
\end{equation}
where $d$ is the element-wise distance of every voxel in the first set to the closest voxel of the same label in the second, $\hat{Y}$ are the predicted labels of each voxel, $Y$ are the ground truth labels of each voxel, and $||$ is the concatenation operator.

\subsection{Performance}\label{sec:performance}

\begin{table}[h]
\caption{Mean results on validation set.}
\begin{center}
\begin{tabular}{| l | c | c | c | c | c | c |}
\hline
\cline{2-7}
\multicolumn{1}{|c|}{Metric} &
       \multicolumn{3}{c|}{Dice}  & \multicolumn{3}{c|}{HD95} \\
\hline
\multicolumn{1}{|c|}{Tumor Subregion} &
        WT & TC  & ET &
        WT & TC & ET \\
\hline
GNN
        & 0.874 & 0.782 & \textbf{0.738}
        & 6.92 & 16.67 & \textbf{20.40}\\
GNN-CNN
    & \textbf{0.894} & \textbf{0.807} & 0.734
    & \textbf{6.79} & \textbf{12.62} & 28.20\\

\hline
\end{tabular}
\label{tab:mean_results}
\end{center}
\end{table}

\begin{table}[h]
\caption{Median results on validation set.}
\begin{center}
\begin{tabular}{| l | c | c | c | c | c | c |}
\hline
\cline{2-7}
\multicolumn{1}{|c|}{Metric} &
       \multicolumn{3}{c|}{Dice}  & \multicolumn{3}{c|}{HD95} \\
\hline
\multicolumn{1}{|c|}{Tumor Subregion} &
        WT & TC  & ET &
        WT & TC & ET \\
\hline
GNN
        & 0.906 & 0.885 & 0.813
        & 3.46 & 3.16 & 2.45\\
GNN-CNN
    & \textbf{0.925} & \textbf{0.908} & \textbf{0.842}
    & \textbf{3.00} & \textbf{3.00} & \textbf{2.24}\\

\hline
\end{tabular}
\label{tab:median_results}
\end{center}
\end{table}

The mean and median results on the validation set are given in Tables \ref{tab:mean_results} and \ref{tab:median_results}, respectively. On the validation set, we report both the performance of the GNN model and of the joint GNN-CNN model. 

The comparison of the two models shows that the addition of the convolutional layers to the model improves mean and median performance across both metrics in the whole tumor and core tumor regions, and is inconclusive for the enhancing tumor. In the case of ET, the CNN improves the average segmentation (better median), but also seems to exacerbate poor performance on outliers (worse mean). Nonetheless, the overall improved results indicate that the addition of the CNN can successfully correct misclassification errors that result from mixed-label supervoxels, even while the CNN architecture is very simple. Notably, the median improvement across all three subregions demonstrates that the joint GNN-CNN model is 1) better able to distinguish the border edema from healthy tissue, 2) better able to distinguish NET from edema, and 3) better able to distinguish ET from NET on a typical brain.

\begin{figure}[ht]
\centering
\includegraphics[width=0.6\textwidth]{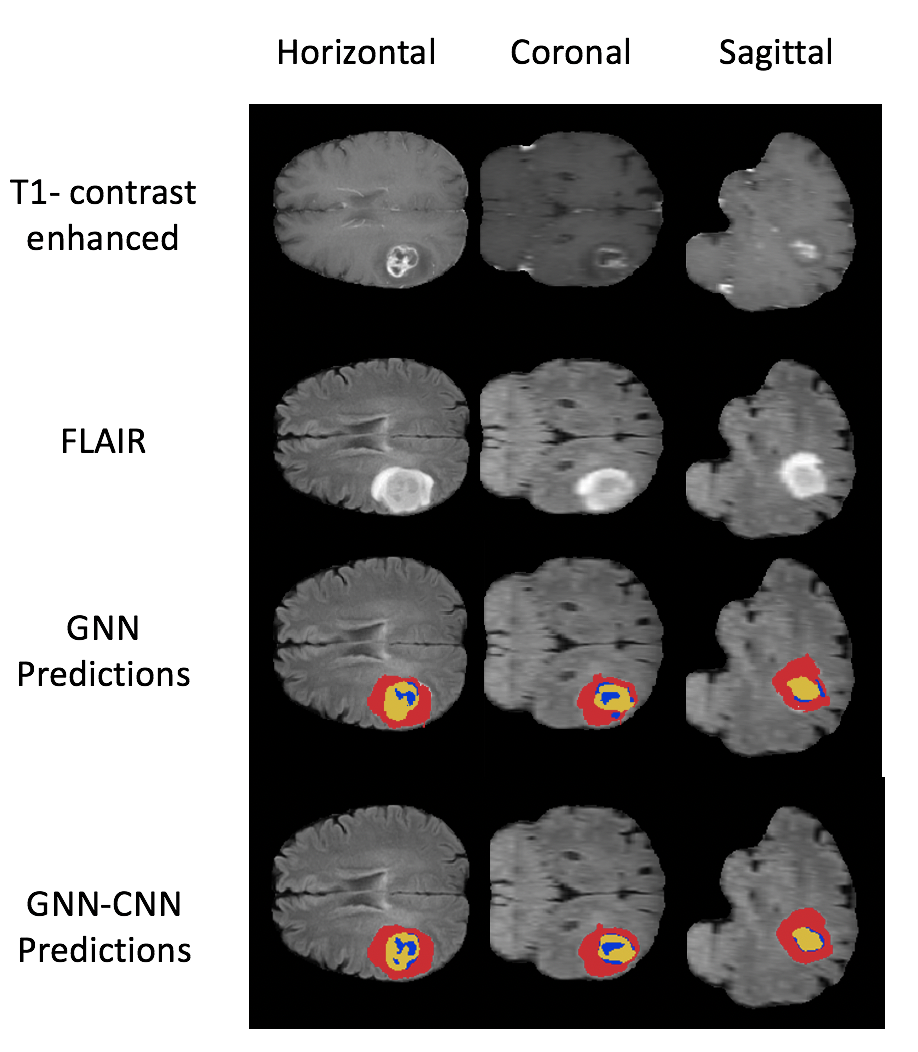}
\caption{\textbf{Example Predictions on Validation Brain} Three slices (horizontal, coronal, and sagittal) of the same brain from the validation set are shown. The first row is from the T1ce modality, and the second is from the FLAIR modality. The third shows the GNN predictions. The fourth row contains the GNN predictions refined through the CNN. Ground truth segmentations are unavailable for the validation set. Red = edema, Blue = NET/necrosis, Yellow = ET. We observe that the GNN accurately identifies the tumorous region but makes slight errors in classifying the individual compartments. The CNN, however,  can refine the predictions in greater accordance with the images.} \label{ex_results_val}
\end{figure}

An example segmentation highlighting these improvements is provided in Figure ~\ref{ex_results_val}, along with two of the four input modalities. The FLAIR image provides information on the tumor core and edema and is thus well suited for the segmentation of the whole tumor. The T1ce modality provides complementary information on NET/necrotic tissue and the enhancing tumor and is thus vital for delineation of the ET and NET subregions. The predictions that have been refined through the CNN (last row) are both smoother and correspond more closely with the shape and appearance of the tumor in the two modalities than the predictions made directly by the GNN (third row).

Given its superior performance on the validation set, we chose the joint model for evaluation on the test set. These results are provided in Table~\ref{tab:test_results}. The test set consists of 570 images. Of these, 87 have a different orientation than the images in the train and validation set. Unfortunately, the challenge organizers informed us that our model submission was unable to produce segmentations for these 87 images. Nonetheless, to preserve consistency across all participants, they have been included in the aggregated results with Dice scores of 0 and Hausdorff distances of 300.

\begin{table}[h]
\caption{Results on test set.}
\begin{center}
\begin{tabular}{| l | c | c | c | c | c | c |}
\hline
\cline{2-7}
\multicolumn{1}{|c|}{Metric} &
       \multicolumn{3}{c|}{Dice}  & \multicolumn{3}{c|}{HD95} \\
\hline
\multicolumn{1}{|c|}{Tumor Subregion} &
        WT & TC  & ET &
        WT & TC & ET \\
\hline
Mean
        & 0.747 & 0.680 & 0.560
        & 63.15 & 72.63 & 75.74\\
Median
    & 0.911 & 0.884 & 0.703
    & 3.74 & 3.16 & 3.31 \\

\hline
\end{tabular}
\label{tab:test_results}
\end{center}
\end{table}

On the test set, the median results approach those achieved on the validation set, but the mean scores fall far below the expected performance. We suspect that the discrepancy between mean and median scores is caused by the inclusion of the 87 failed cases. The existence of such outliers would skew the mean more than the median scores, leading to the observed pattern. Nonetheless, the median results indicate that, on a typical unseen tumor, our model is effective at locating the whole and core tumor, but has difficulty delineating the enhancing tumor from surrounding regions. Possible improvements to ET prediction are considered in the discussion.



\section{Discussion}

We have presented a joint GNN-CNN network for automatic brain tumor segmentation. The GNN can produce good segmentations on its own, but struggles to accurately delineate exact tumor and tumor compartment boundaries due to the coarse supervoxel generation step. We show that this limitation can be at least partially circumvented by adding convolutional layers to the end of the model to smooth out predictions. While it is likely that a more complex CNN could further boost performance, this work aims to improve the feasibility of GNNs for tumor segmentation rather than to engineer an optimal CNN.

A clear direction for future work is to diagnose the failure cases of our model. In particular, our model should be able to produce a segmentation on any volume, regardless of orientation. It is likely that this issue is technical rather than a failure of the model to generalize, but it is difficult to identify without access to the testing data. Furthermore, it will be interesting to explore how segmentation of the enhancing tumor can be improved. The enhancing tumor is typically a small or set of small regions, which makes it inherently harder to accurately delineate with supervoxels. Perhaps a hierarchical segmentation scheme or more complex CNN will be able to improve model performance here. It has also been demonstrated by other participants of this year's challenge that post-processing heuristics to remove false positive ET predictions can have a meaningful impact on performance. Lastly, we also aim to incorporate a soft Dice loss in future work to improve the predictions of the composite tumor regions, rather than just the individual subtypes.



\section{Code availability}

Our code is publicly available at \url{https://github.com/rsinghlab/GNN-Tumor-Seg}

%
%
\bibliographystyle{splncs04}
\bibliography{refs}

\end{document}